# Building Reliable Arithmetic Multipliers Under NBTI Aging and Process Variations

Masoud Heidary, Biresh Kumar Joardar, Department of ECE, University of Houston

*Abstract*— **Hardware aging poses a significant challenge for integrated circuits (ICs), leading to performance degradation and eventual failure. In this work, we focus on the aging of arithmetic multipliers, which are a cornerstone of modern computing systems including in CPUs, GPUs, and FPGAs, as well as AI accelerators like systolic arrays. In particular, AI workloads, which rely predominantly on multiplications, can accelerate Negative Bias Temperature Instability (NBTI) effects in multipliers. This paper presents a novel aging mitigation technique that leverages the sign-invariance property of multiplication (A × B = (–A) × (–B)). By selectively applying 2's complement transformations to inputs, the method redistributes stress across transistors, reducing the effects of NBTI aging. The proposed method is also integrated into systolic arrays, a common AI accelerator, to demonstrate its efficiency in a high-throughput AI accelerator. Experimental evaluations using Cadence tools show up to 75% better lifetime compared to natural aging (with no mitigation) baseline, while introducing negligible area and delay overheads.**

## I. INTRODUCTION

As semiconductor technology advances toward smaller nodes, hardware aging has emerged as a significant challenge in modern computing systems [1]. Aging mechanisms such as Bias Temperature Instability (BTI) [2], Hot Carrier Injection (HCI) [3], and Time-Dependent Dielectric Breakdown (TDDB) [4] gradually degrade transistor performance, leading to reduced speed, higher power consumption and even failure [5][6]. In this work, we focus on negative bias temperature instability (NBTI), which is a critical aging mechanism in any IC, including systolic arrays [7]. NBTI predominantly impacts PMOS transistors that experience prolonged stress due to negative gate bias. Following convention [8], *we define a PMOS is under stress when gate input is logic '0'*. Over time, NBTI leads to a gradual increase in threshold voltage ($V_{th}$) [9]. As $V_{th}$ rises, the transistor requires more time to switch, introducing subtle timing shifts that accumulate across cycles. In complex systems like systolic arrays, timing variations can disrupt data propagation, leading to synchronization issues [10] and ultimately diminishing the efficiency and reliability of the multiplier, and thereby the overall device.

Here, we specifically focus on the effects of NBTI aging in arithmetic multipliers, due to their widespread use in ALUs of CPU, GPU and in AI accelerators like systolic arrays, making it one of the most common and ubiquitous hardware blocks [11], [12]. Compute intensive tasks such as AI, which predominantly rely on matrix multiplications, also result in high utilization of the multipliers making them vulnerable to aging effects. In fact, prior work [13] found that multipliers are susceptible to aging more often than other arithmetic components. This observation is also supported by our findings. Figure 1 compares the stress distribution (measured as probability that logic is '0') across different PMOS in a generic array multiplier and ripple carry adder obtained via Cadence simulations. As we can see from Figure 1(a) and Figure 1(b), the distribution of stress is very uniform for ripple adders but varies widely for array multipliers. This results in unequal amount of stress on the transistors in the multiplier. The high stress due to NBTI in some PMOS cause device wear out, gradually degrading performance and reliability over time, leading to erroneous computations [14].

The problem is further exacerbated in AI accelerators such as systolic arrays, where there is unequal usage of the processing elements, often due to non-uniform workload [15], [16]. Process variation and input data patterns also cause certain transistors to age faster than others. Such unequal aging causes timing mismatches among various paths in the hardware, resulting in computation inaccuracies, and eventual device failure, which directly impacts tasks supported by these architectures such as AI inferencing [17], [18]. Hence, we focus on improving the reliability of multipliers here. We discuss how the idea can be extended to other arithmetic operations and hardware units later.

To mitigate the effects of aging, hardware designers have explored various techniques such as dynamic voltage and frequency scaling (DVFS) [19], material level optimizations [20], thermal aware system design [21], etc. These approaches aim to reduce transistors stress by adjusting operating conditions or improving manufacturing processes. Another approach involves workload balancing, where computations are distributed across different PEs to even out the stress [16]. However, these existing solutions often introduce additional and often unwanted software changes, require external intervention, or limit the overall throughput of the hardware making them inefficient for high performance tasks like AI acceleration.

In this work, we propose a simple hardware-level aging mitigation technique that leverages the mathematical property:

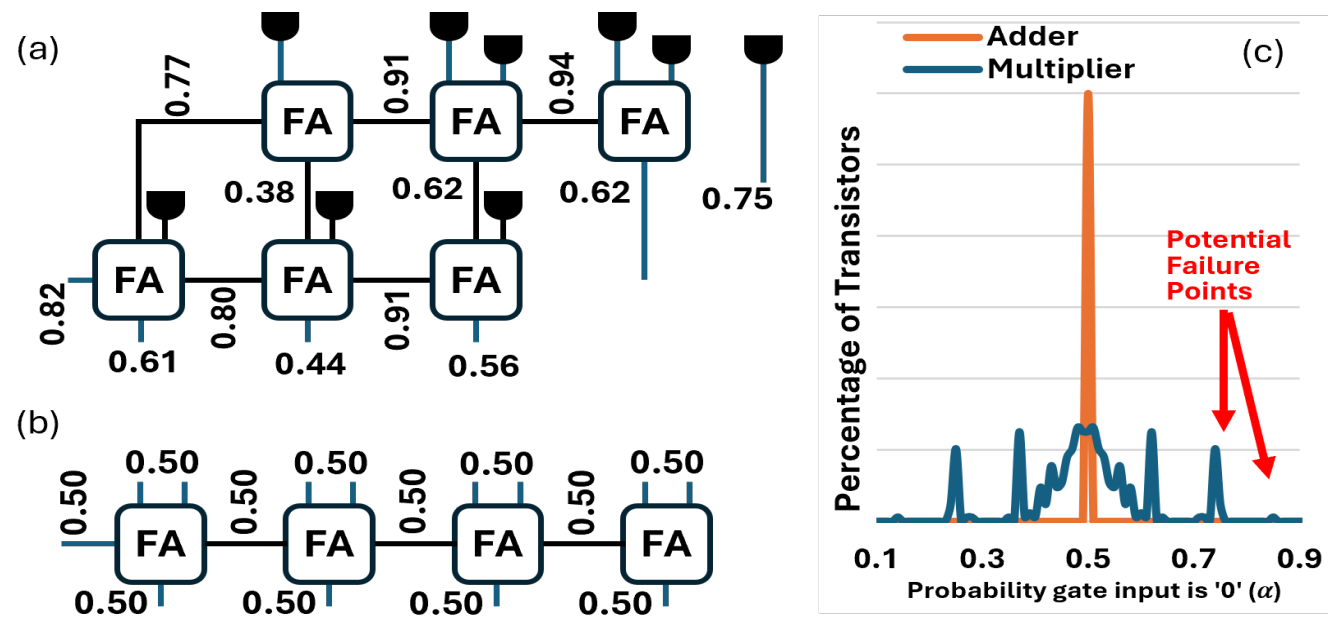

**Figure 1.** Probability that input is '0' ($\alpha$), indicating regions with potential stress in (a) ripple adders, and (b) array multipliers, and (c) distribution of $\alpha$ across all PMOS in adder and multiplier.

$-A \times -B \equiv A \times B$ i.e., multiplications using 2's complement yields same results as normal representation. While the computed results are identical, our analysis shows that the resulting stress distributions differs significantly between the two encodings. By selectively applying 2's complement transformations at runtime, the proposed method redistributes stress across the multiplier's transistors, reducing localized degradation caused by aging, input randomness and process variation. By identifying patterns in the input data, we first develop an intelligent way to decide when to apply 2's complement transformation to the inputs. By selectively applying the transformation, the proposed method changes the probability of stress on the PMOS transistors and extend the operational lifetime of the hardware.

The proposed technique offers several advantages over traditional aging mitigation methods. First, it can be implemented using dedicated hardware with little additional area, eliminating the need for any software modifications or complex runtime scheduling. Second, it requires minimal changes to the existing architecture, allowing easy integration. Third, due to the ubiquitous nature of multipliers, the proposed method can benefit various architecture including CPU, GPU and AI accelerators like systolic arrays, FPGA, etc. Experimental analysis show that our approach achieves significant lifetime improvements without degrading computational performance, making it an attractive solution for extending the reliability of various architectures. Through systematic analysis and experimental validation, we demonstrate that the proposed method achieves up to 75% lifetime improvement in systolic array-based AI accelerators.

The rest of the paper is organized as follows. Section II provides background on NBTI aging and process variation. Section III discusses prior works. Section IV introduces the proposed method, including the design of a Selector Module (SM) and its integration into systolic array. Section V presents our experimental setup, evaluation, and baseline comparisons to assess the effectiveness of our proposed method. Finally, Section VI concludes the paper.

## II. BACKGROUND

In this section, we present a background discussion on NBTI aging and process variation.

### A. NBTI aging

As mentioned earlier, NBTI leads to performance degradation, causing an increase in threshold voltage ($V_{th}$) over time ($t$), which is given by the following equation:

$$\mathrm{V_{th}}(t) = V_{th}(t=0) + |\Delta V_{th}(t)| \quad (1)$$

The change in $V_{th}$ at time $t$ (denoted as $|\Delta V_{th}(t)|$) results in a decrease in current, and an increase in circuit's delay. This degradation negatively impacts the reliability of digital circuits, particularly in high-performance AI accelerators where precise timing and consistent transistor performance are essential [22].

NBTI aging occurs primarily, when a PMOS transistor is in the ON state ($V_{gs} = -V_{dd}$). In this state, holes from the transistor's inversion layer interact with silicon-dioxide ($SiO_2$) defects, leading to the generation of interface traps at the $Si - SiO_2$ boundary. These traps capture charge carriers, increasing threshold voltage and slowing down the transistor's switching speed. The analytical model for long-term NBTI degradation is given by the Reaction-Diffusion model [8]:

$$|\Delta \mathrm{V_{th}}(t=0)| = \left(\sqrt{K_v^2.\alpha.T_{data}}\,/\left(1-\beta(\mathrm{t})^{1/2\lambda}\right)\right)^{2\lambda} \quad (2)$$

Here, $T_{data}$ is the period of time an input is applied, $\alpha$ is the probability that the input at the PMOS gate is '0', resulting in negative bias at a PMOS, and $\lambda$ represents the time exponent. Overall, the term $\alpha.T_{data}$ represents the duration when a PMOS is under stress during use. For more details about the model and relevant parameters, we refer the reader to [8], [23]. Together, Equation (1) and Equation (2), represent the behavior of $V_{th}$ in a PMOS transistor over time. Overall, as $V_{th}$ increases, the transistors within the multipliers and adders exhibit slower switching speed, which leads to increased delay in multiply and accumulate operations and higher power consumption. In the context of AI accelerators, this can lead to lower model accuracy, as increased delay can cause timing violations.

### B. Process variation

Process variation is a challenge in IC manufacturing, leading to deviations in transistor characteristics across a chip. These variations arise from imperfections in fabrication process such as lithography, doping, etching, and oxide growth. These variations lead each transistor to have slightly different properties (e.g., different $V_{th}$), leading to inconsistencies in circuit performance [24]. Transistors with an initially higher threshold voltage may experience accelerated degradation due to NBTI. Unequal aging can result in some transistors failing sooner than others, resulting in unbalanced paths, and more errors. The impact of process variation on $V_{th}$ (referred as $\Delta V_{th,PV}$) can be mathematically expressed using a Gaussian distribution denoted as $N$ [25]:

$$\Delta V_{th,PV} \sim N\left(0, \sigma_{V_{th}}^2\right) \quad (3)$$

The initial threshold voltage at time $t = 0$, is influenced by both nominal threshold voltage ($V_{th,nominal}$) and frabrication process variation ($\Delta V_{th,PV}$), expressed as below:

$$\mathrm{V_{th}}(\mathrm{t}=0) = \mathrm{V_{th,nominal}} \pm \Delta V_{th,PV} \quad (4)$$

The nominal threshold voltage $\left(V_{th,nominal}\right)$ refers to the threshold voltage specified by the manufacturer. Overall, the threshold voltage at time $t$ $\left(V_{th}(t)\right)$, considering both process variation and NBTI aging, can be obtained by substituting Equation (4) and Equation (2) in Equation (1):

$$\mathrm{V_{th}}(\mathrm{t}) = \mathrm{V_{th,nominal}} \pm \Delta V_{th,PV} + |\Delta \mathrm{V_{th}}(t)| \quad (5)$$

We use this equation to estimate the threshold voltage, which is then incorporated in circuit-level simulations (e.g., Cadence Ocean) to obtain performance, power, etc.

## III. PRIOR WORK

Hardware aging is a critical challenge in modern computing systems, affecting both reliability and performance over time [1]. Various aging mechanisms degrade transistor characteristics, leading to failure or increased power consumption. Among these, NBTI, HCI, and TDDB are the most prominent[3], [7], [26]. NBTI primarily affects PMOS transistors, increasing their threshold voltage ($V_{th}$) over time, which in turn increases delay and reduces drain current, leading to potential computational errors [7]. HCI impacts both NMOS and PMOS transistors, causing degradation due to high-energy carriers trapped in the oxide layer, which leads to increased resistance and reduced transistor performance [3]. TDDB, on the other hand, occurs when the gate oxide gradually breaks down, leading to leakage currents and eventual failure [26]. The degradation due to aging is further exacerbated by process variation. Due to variations from manufacturing and usage, some transistors age at a faster rate than the others [27]. The non-uniform aging induces varying amounts of slowdown in different paths, making it a challenging problem to address.

To compensate for these aging effects, researchers have developed various strategies aimed to prolong the operational lifetime of hardware components. Many prior works have contributed valuable insights into aging mitigation, focusing on techniques such as voltage scaling [19], error correction [28], workload balancing [29], instruction set encoding [30], and process variation-aware strategies [16]. The Dynamic Voltage and Frequency Scaling (DVFS) method dynamically adjusts voltage and frequency based on workload demand. Lowering voltage during low activity period reduces transistors stress, slowing down the aging process [19]. While DVFS is effective, it relies on the assumption that period of low activity can be determined in real-time. Moreover, it introduces performance overhead, as lower voltage leads to decreased throughput.

Error correction and reliability-aware scheduling are other common strategies [28], [29]. However, error correcting techniques are limited to correcting one (or few) errors only. In reliability-aware scheduling, tasks are dynamically reassigned to distribute workload to avoid excessively aging some hardware units. However, these approaches rely on software-level optimization and complex scheduling policies, which do not directly reduce transistor stress. Aging-aware instruction set encoding has been proposed to minimize stress on transistors [30]. While this is an efficient way to mitigate aging in processors, it does not apply to systolic arrays, where computations are less diverse.

Workload distribution based on process variation was proposed in [16]. By assigning lower-stress tasks to processing elements (PEs) that show signs of accelerated aging, this method reduces the impact of process variation on system reliability. However, it relies on the assumption that the systolic array processes matrices of varying sizes. If the matrix size is uniform, this method is unable to reduce aging. Moreover, workload distribution introduces scheduling complexity and may reduce throughput due to frequent reallocation of tasks. Another method, referred as dTune, introduces a process variation aware compiler to mitigate aging by optimizing the code execution pattern [31]. However, this method lacks real-time condition about hardware and may result in sub-optimal outcomes if aging distribution changes over time.

In this work, we advance the state-of-the-art by proposing a new aging mitigation approach. The proposed approach takes advantage of the inherent properties of multiplications, and it can be implemented with dedicated hardware, with negligible performance, energy and area overheads as we show later. The proposed method directly reduces transistor stress by dynamically shifting stress away from heavily degraded transistors in multipliers and systolic arrays. Experimental results show significant improvement in device lifetime with negligible area, power and performance overheads.

## IV. PROPOSED METHOD

In this section, we present the proposed technique in detail.

### A. Overview

While various methods exist to reduce/slow aging, here we propose a *2's complement transformation-based approach* that dynamically redistributes transistor stress to extend hardware lifetime. The core idea leverages the sign-invariance property of multiplication ($A \times B = (-A) \times (-B)$) to selectively apply 2's complement transformations to input operands. While this transformation might seem trivial, it can redistribute stress among transistors, without affecting output correctness.

Figure 2 demonstrates how the proposed idea changes stress among transistors in a multiplier. Specifically, the proposed method shifts stress from transistors that have aged significantly, to other less-aged transistors using 2's complement. As we can see in Figure 2(a), multiplying $100 \times 1$ in a 8-bit array multiplier causes the transistors marked in red (including T1) to be under stress (since gate input is '0'). Next, if we multiply $-100 \times -1$, in Figure 2(b), the final output remains unchanged. However, some of the PMOS gate inputs change. For instance, the gate input of T1 changes from logic $0 \rightarrow 1$, thereby reducing stress on T1.

The above example is how applying 2's complement transformation to input can redistribute stress among the various transistors of multiplier, thereby affecting the overall lifetime of the circuit. Existing literature commonly defines the end-of-life (i.e., lifetime) of a MOSFET as the point at which its saturation current is reduced by 10-25% relative to a new device [32], [33], [34]. In this work, we assume a more conservative limit of 50%, i.e. *we define "lifetime" as the point in time when the ON current of at least one transistor in the multiplier drops to 50% of its unaged counterpart*. Please note that this threshold is chosen for illustration; the proposed method applies to any current-based definition of lifetime.

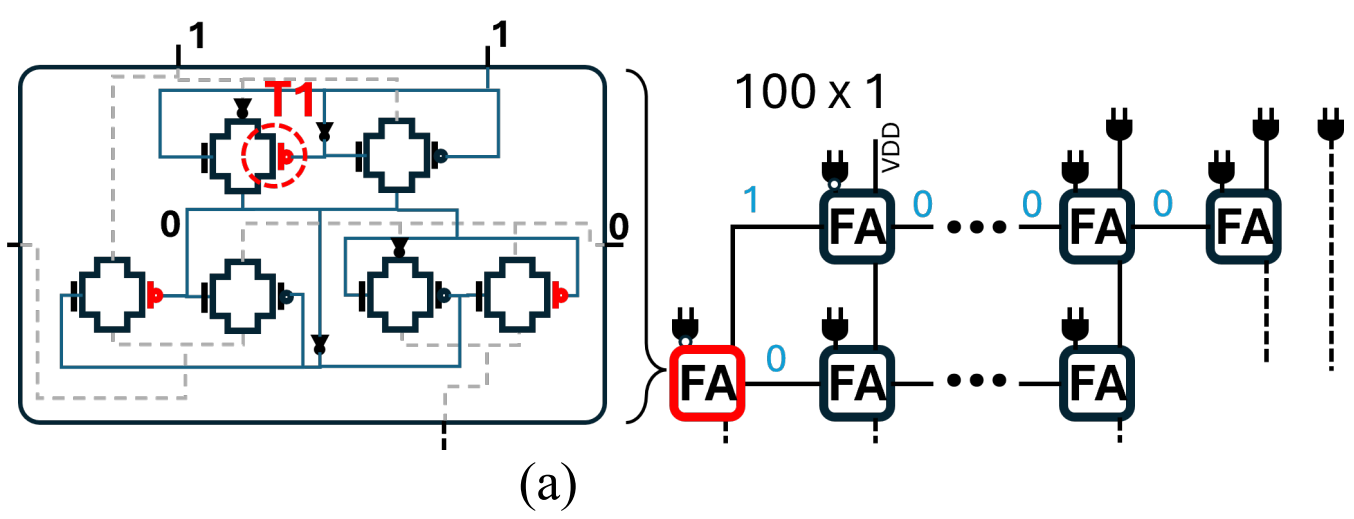

(a)

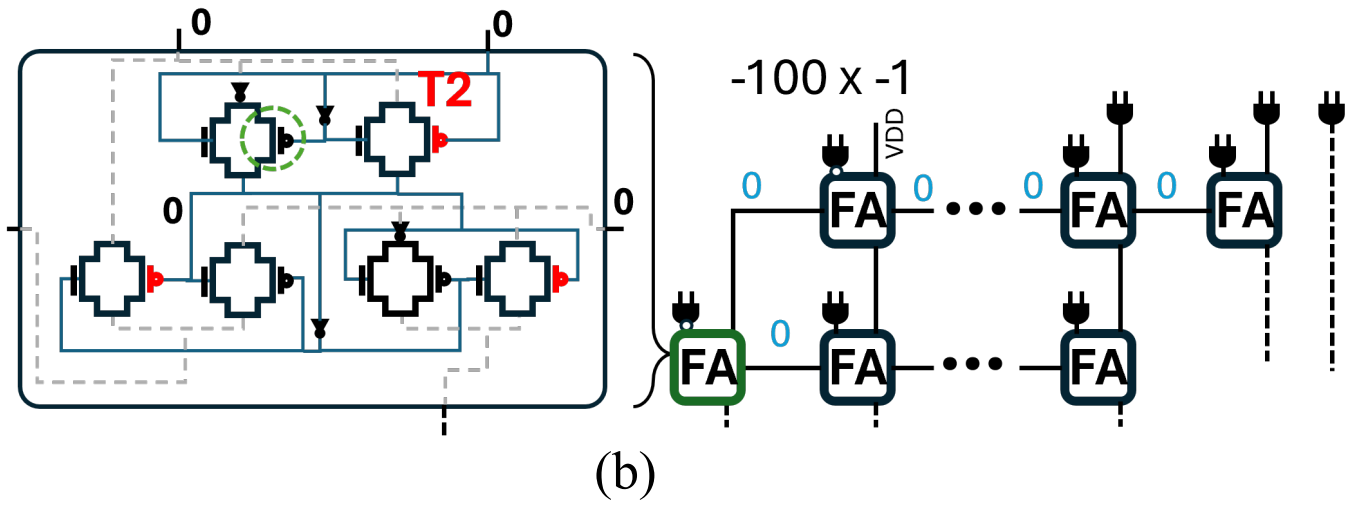

(b)

**Figure 2.** Illustration of how the use of 2's complement can redistribute stress in an array multiplier when (a) multiplying $100 \times 1$, and (b) its 2's complement version $-100 \times -1$. As we can see here, stress on T1 reduces when multiplying $-100 \times -1$.

Despite its simplicity, using 2's complement transformation to mitigate aging faces a critical challenge. *The transformation must be applied selectively depending on the aging condition of the various transistors in the circuit (multiplier here).* Applying the transformation to all inputs *indiscriminately* does not result in improvement in the lifetime of the hardware. For instance, in Figure 2, instead of T1, if T2 has experienced accelerated aging, then applying the same transformation $100 \times 1 \rightarrow -100 \times -1$, can lead to higher stress on T2, thereby degrading T2 even further, resulting in lower lifetime. Hence, applying the 2's complement transformation must consider the amount of degradation of the various PMOS in the multiplier.

Overall, we must be selective in applying the transformation, considering which transistors are degraded most. By carefully selecting when and where to apply this transformation, we can shift stress away from heavily degraded transistors, balancing wear out across the various PMOS. Unlike traditional aging mitigation techniques that require complex scheduling or architectural changes, the proposed method introduces little hardware overheads and can be easily implemented in the arithmetic units as we show next.

### *B. Creating the ideal selector (Oracle)*

While 2's complement approach can change stress distribution, it is important to first identify *which inputs to transform* considering the most degraded/stressed transistors. However, determining which subset of inputs to transform is not straightforward. First, aging is affected due to process variation in ICs. Multipliers in different ICs will have different sets of transistors that experience accelerated aging. Hence, which inputs to transform must consider aging imbalance due to process variation and optimized on a case-by-case basis. Moreover, the selection process must consider the fact that the proposed method shifts stress from one PMOS to another PMOS in the multiplier. For instance, in Figure 2, by applying 2's complement, there is reduction in stress on T1, but it happens at the cost of increased stress on T2, which can become the new bottleneck. Hence, we first aim to develop an ideal selector (Oracle) that can predict when to apply 2's complement transformation for a given input such that the overall stress is reduced, and lifetime is maximized.

For this purpose, we develop a search process to identify when to apply the transformation based on the most likely 'First-to-Fail' (F2F) transistor(s) (Line 2) as shown in Algorithm-1. As mentioned earlier, we define a transistor as failed when its drain current decreases by 50% compared to its initial value. We start by identifying the condition of various transistors in the multiplier at $t = 0$. We assume that this step is done offline as part of the manufacturing test or in the field via built-in-self-test (BIST). This data is incorporated as part of the NBTI model (Equation (2)) to estimate $V_{th}$, which is then used in Cadence simulations to predict PMOS that are likely to fail early i.e., $\mathbb{T}_{F2F} = \{PMOS_i\}$ (Line 9); note that these F2F transistors may not always lie on the critical path. For ease of demonstration, we first assume that there is no process variation in Algorithm-1. We extend the algorithm to include process variations later.

As shown in Algorithm-1 (Line 4), we iterate through every input $I = \{a, b\} \in (A, B)$. For each input, we note the stress, on F2F transistors ($\mathbb{T}_{F2F}$) considering both $I = \{a, b\}$ and its 2's complement $I' = \{-a, -b\}$. If the overall stress on F2F transistors is reduced with $I'$, we mark it as '$True$' in the Oracle i.e., $\varphi(I, I') = True$. Estimating stress (Line 5) can be done using a circuit simulator (e.g., Cadence Virtuoso/Ocean). However, using $I'$ (instead of $I$) sometimes result in a non-F2F transistor becoming a new bottleneck as stress is redistributed. To prevent such scenarios, we note the new F2F transistors (if any) denoted as $\mathbb{T}'_{F2F}$ in Line 7, when $I \rightarrow I'$ transformation is applied. If no new F2F transistor are found i.e., $\mathbb{T}'_{F2F} \subseteq \mathbb{T}_{F2F}$, we stop, else, the process continues iteratively with the updated list of F2F transistors as shown in Line 8.

We repeat this process for all inputs $I = \{a, b\} \in (A, B)$, to create the ideal selector/Oracle, denoted as $\varphi(I, I')$. At the end, the Oracle $\varphi(I, I')$ can be envisioned as a table that stores binary data denoting whether to apply transformation to an input (or not) for a given set of F2F transistors, such that the multiplier's lifetime is increased. Finally, it should be noted that oracle creation is a *one-time offline process* for a given multiplier architecture, and thus, we do not consider the time required here as an overhead. Moreover, since a multiplier is a relatively small circuit, the Oracle creation can be done easily using circuit simulation tools like Cadence Virtuoso/Ocean.

### *C. Hardware-friendly selector module*

Next, we propose to implement the Oracle's behavior in hardware via a selector module (SM) that can guide whether to apply the transformation (or not), in real-time during use. While the Oracle $\varphi(I, I')$ can be implemented as a lookup table, it is not practical due to its high overhead. For an 8-bit multiplier,

---

**Algorithm 1** Generating Ideal Selector

1: **getOracle**():
2: $\quad \mathbb{T}_{F2F} = \boldsymbol{getF2Ftransistor}(\mathbb{V}_{th,nominal}, \mathbb{P}_{nominal})$
3: $\quad$ while($True$):
4: $\quad\quad$ FOR $inputs\ I_i \in (\mathbb{A}, \mathbb{B})$:
5: $\quad\quad\quad \varphi[I_i] = \begin{cases} True, if\ \boldsymbol{stress}(-I_i, T_{F2F}) < \boldsymbol{stress}(I_i, T_{F2F}) \\ False, otherwise \end{cases}$
6: $\quad\quad \mathbb{P}' = update\ using\ \mathbb{L}$
7: $\quad\quad \mathbb{T}'_{F2F} = \boldsymbol{getF2Ftransistor}(\mathbb{V}_{th,nominal}, \mathbb{P}')$
8: $\quad\quad$ if ($\mathbb{T}'_{F2F} \subseteq \mathbb{T}_{F2F}$): $return\ \varphi$
$\quad\quad$ else: $\mathbb{T}_{F2F} \leftarrow \mathbb{T}_{F2F} \cup \mathbb{T}'_{F2F}$

9: $\boldsymbol{getF2Ftransistor}(\mathbb{V}_{th,t=0}, \mathbb{P})$:
$\quad$ // Simulate aging with NBTI model (Equation (2)) and using Cadence

---

$\mathbb{P}$: Multiplier operating parameters (e.g. $V_{dd}$, $\alpha$, …)
$\mathbb{I}_t$: Current through a PMOS at time $t$

with two inputs $(A, B)$, the total number of input combinations is $2^{16}$ (i.e. 65,536). Creating an ideal selector representing the Oracle using look-up tables is impractical due to the area and performance overheads involved. Moreover, the Boolean expression that represents $\varphi(I, I')$ *exactly,* is complex and implementing it using digital logic incurs high overheads.

Instead, we aim to create a hardware-efficient SM that *closely approximates* the Oracle behavior using fewer bits to reduce overhead. For this purpose, we propose to represent and approximate the Oracle $\varphi(I, I')$ via a simplified logical function $f(A, B) \equiv \varphi(I, I')$. The function $f$ takes $k$ bits from the two primary inputs $A$ and $B$, and it produces a 1-bit $(True/False)$ as output, indicating whether 2's complement should be applied or not. However, determining $f$ is a challenging problem as when input bit-width increases, the number of possible bit patterns grows combinatorially. Specifically, the number of ways to select $k$ bits from a total of 16 input bits is $(16, k) = 16!/(k! * (16 - k!))$.

The above-mentioned problem can be formulated as an *approximate logic synthesis (ALS) task*, which can be solved via heuristics, machine learning, graph representations, etc. [35]. In this work, we use a heuristics-based method, due to its simplicity, noting that other methods can also be used. First, we examine all input bit patterns and their correlation with the outcome $(True/False)$. We choose the $top - k$ bits (where $k$ is user's choice) from this list to determine a suitable (approximate) Boolean logic function $(f)$ that closely resembles the Oracle $\varphi(I, I')$. For example, assuming no process variation and $k = 2$, we find that the most significant bit of input $A$ and the least significant bit of input $B$ have the strongest correlation with $\varphi(I, I')$. Hence, we can make the selector module that implements a Boolean function $f(A_{MSB}, B_{LSB})$ to determine when to apply the 2's complement transformation. We repeat this process for various values of $k$, and evaluate the resulting approximate Boolean logic in terms of accuracy (how closely $f$ resembles $\varphi(I, I')$), while also considering overheads (power, performance, and area). Based on our experiments, we find that using $k = 4\ to\ 6$ bits for $f$ results in a good accuracy-overhead trade-offs.

**Aging of SMs:** Finally, we note that the SMs also age and degrade over time. However, the SMs consist of minimal logic with 4-6 bit inputs and are much faster than the multiplier itself. In a typical pipelined architecture, the delay and operating frequency is limited by the slowest stage (multiplier here). Hence, *even though the SMs get slower as they age, it does not become a performance bottleneck*. Hence, no aging mitigation is required for the SMs.

*D. Process variation and its impact on multipliers*

Process variation can result in up to ~10% deviation in nominal $V_{th}$ [36]. If the $V_{th}$ of a transistor is higher than others, it can fail sooner as it experiences accelerated degradation. Such difference in aging introduces an additional challenge as process variation can result in different sets of F2F transistors. A single SM may not suffice in presence of process variation.

Interestingly, we observe that the proposed method is still applicable despite process variations. To verify this, we did Monte Carlo simulations, where we assume random process variation across the various transistors. In each iteration, we start with a new set of F2F transistors by varying the nominal threshold voltage. We observe that the selector function $f(A, B) \equiv \varphi(I, I')$ (discussed in previous section) can reduce stress and slow aging in most cases (as we show later). However, for some process variation scenarios, we observe no benefit of using just one selector function. To address this problem, we propose to introduce an ensemble of selectors to achieve better accuracy for various process variation scenarios.

The overall idea is similar to the use of ensembles in machine learning literature, where multiple learners are used together to get better predictions [37]. Hence, we explore the use of multiple SMs to reduce stress under different process variation scenarios here. The additional selectors (referred as $SM - i$) can be developed following the same process (Algorithm-1) considering different F2F transistors. The value of $i$ is decided based on lifetime-area overhead trade-off desired. At the end, we have multiple selector modules, where each module is based on a different Boolean function. Depending on which transistor(s) is/are F2F, we choose the most appropriate selector. The chosen SM is then responsible for applying the 2's complement transformations. From our experiments, we found that using 2-4 different SM is sufficient even in presence of process variation. Using more selectors leads to no significant gains in lifetime as we show later.

*E. Final implementation considering an AI accelerator*

Finally, we demonstrate the effectiveness of proposed setup within a systolic array-based AI accelerator to showcase its integration in a realistic and high-performance computing context. Figure 3 shows the final architecture.

Systolic arrays are highly parallel, structured architectures commonly used to accelerate matrix-based operations like Multiply-Accumulate (MAC), which form the core of many scientific, signal processing, and machine learning workloads. A systolic array (SA) consists of many PEs, each PE includes an adder and a multiplier. By enabling data to move between adjacent PEs, systolic arrays achieve high throughput and energy efficiency while minimizing memory access [16].

A typical SA consists of an $D \times D$ grid of PEs, each responsible for performing MAC operations. During matrix multiplication, input matrix $X$ is streamed in word-line across rows, while matrix $W$ is streamed in bit-line down the columns. Each PE receives a pair $(X_r, W_c)$, performs a multiplication, and adds the result to a running sum stored in an internal register.

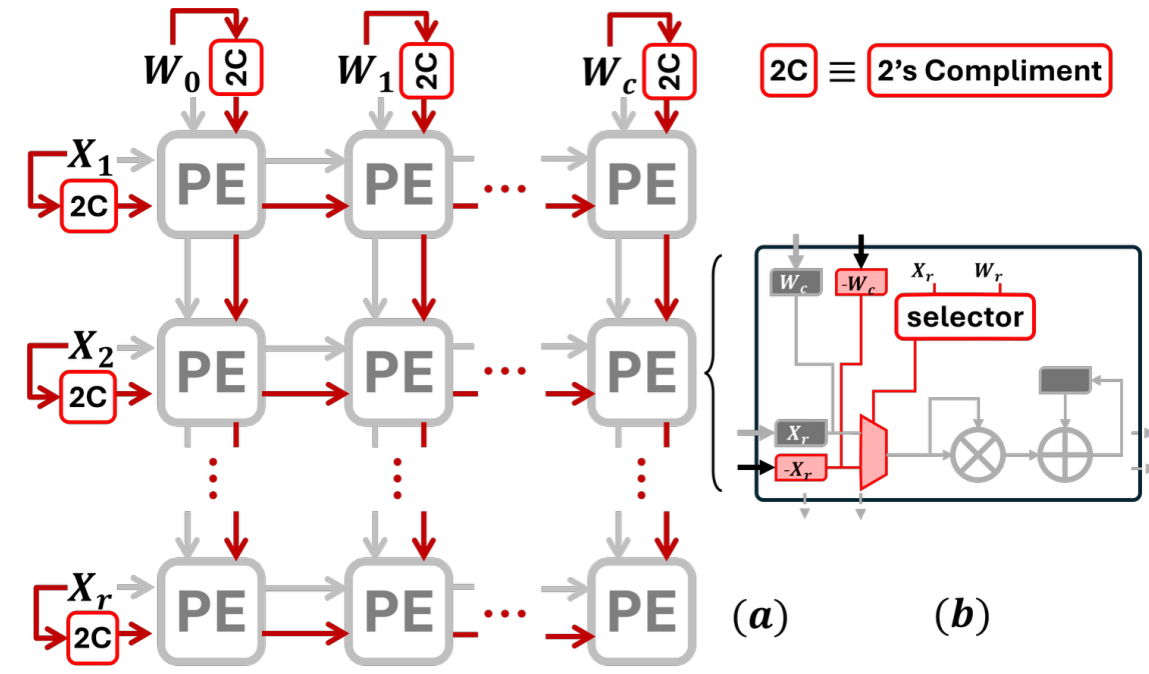

**Figure 3.** A typical (a) Systolic array architecture and (b) processing element (PE) design. Both have been modified to incorporate the proposed method (new addition are marked in red in the Figure).

To incorporate the proposed aging mitigation strategy within this setup, we modify the architecture by integrating an SM in each PE of the SA architecture (marked in red in Figure 3(b)). This is done as the various PEs in the SA may undergo different amounts of aging, resulting in unequal stress distributions across the various PMOS. Hence, we add SMs in each PE to cater to each PE's specific aging condition. While the SMs are added to each PE, we assume that the 2's complement is pre-calculated only once at the beginning of each row and column. Both inputs ($I$ and $-I$) are then routed to the PEs. The SM in each PE then chooses the required input ($I$ or $-I$) based on its aging condition and F2F transistors.

The SM itself is implemented using CMOS-based digital logic with 4-6 bits input representing the approximation of the ideal selector / Oracle as discussed in Section IV. The use of an approximate function allows us to simplify the Boolean expression of the SM, which we implement using few simple gates (e.g., NAND, NOR) to produce a binary $True/False$ output, determining whether applying the 2's complement will reduce aging stress on F2F transistors. We refer to the hardware module implementing the selector function as SM. The selection logic significantly reduces stress and stress imbalance over time while introducing little power, performance and area overheads as we show next.

## V. EXPERIMENTAL RESULT

In this section, we present the experimental evaluation of our proposed lifetime aware selector architecture.

### *A. Experimental Setup*

**Hardware setup:** To evaluate the proposed approach, we assume a typical 128×128 SA architecture, noting that other dimensions can also be used. Each PE in the SA consists of an adder and multiplier along with other peripherals as outlined in Figure 3. For thorough experiments, we have assumed 4, 8, 12, 16, and 32-bit fixed-point multipliers, noting that it can be applied to floating-point operations as well (as we discuss later). We also consider two multiplier architectures, namely the Array and Wallace Tree Multiplier to thoroughly evaluate the proposed method across various bit widths and architecture. For hardware-level evaluation, we used a mix of Cadence Virtuoso (for designing the circuit) and OCEAN (for simulation). The simulation was conducted using GlobalFoundries' 22nm (GF22nm) process technology. The supply voltage ($V_{dd}$) in GF22nm is 0.8V, and the threshold voltage of all PMOS at $t = 0$ is 0.45V (without process variation). Other circuit parameters are set as recommended for the GF22nm process.

**Incorporating aging and process variation:** Next, we capture the effect of NBTI aging and process variation on $V_{th}$ using Equations (1)-(5), which is then incorporated in the Cadence simulations to determine the lifetime, area, delay and energy for various configurations, assuming an operating temperature of 25°C. The model parameters for Equations (1)-(5) are obtained from prior work [8], [23]. To model process variation, we assume Gaussian distribution using Equation (4) with standard deviation of $\sigma_{V_{th}} = 0.02$. We perform Monte Carlo simulations with ~300k iterations to thoroughly evaluate the proposed method under various scenarios. In each iteration, we vary the initial threshold voltage $\left(V_{th,}(t = 0)\right)$ of the PMOS transistor randomly following Equation (4).

**Comparison with prior work:** For comparison, we contrast the proposed method to a multiplier that experiences normal aging and has no mitigation technique (baseline). We also compare the proposed method with other aging-aware techniques: DVFS and 1's complement transformation, from prior work [19], [38], [39]. The techniques proposed in [38], [39], both rely on applying 1's complement transformation to mitigate aging randomly. We choose these two techniques as baselines for comparison as they are the closest to the proposed method. The method in [19] uses DVFS to regulate voltage and frequency to reduce circuit stress and aging. The DVFS controller adjusts supply voltage as needed to compensate for any drop in current due to aging, while aiming to increase lifetime. As we show later, the proposed method outperforms all these baselines in terms of lifetime.

### *B. Lifetime improvement using 2's complement*

First, we show evidence demonstrating the efficacy of 2's complement as an effective mechanism to slow NBTI and extend lifetime for a single multiplier. Following Algorithm-1, we first develop the ideal selector (Oracle). The Oracle tells whether we should (or should not) apply the 2's complement transformation to an input. Note that the Oracle represents the best-case and provides the maximum gain in lifetime by precisely applying the transformations when needed. Figure 4 illustrates the distribution of stress in an Array multiplier quantified as the probability that gate input is '0' ($\alpha$), both with and without using the ideal selector. As observed, the use of SM produces two key effects: first, it shifts the overall distribution of $\alpha$ closer to 0.5, thereby reducing the unbalanced aging across the circuit. Second, it lowers the maximum $\alpha$ value present in the design which directly slows the failure of the most vulnerable transistors (i.e., F2F transistors).

Figure 5 shows the expected lifetime because of this distribution shift using the Oracle compared to the baseline scenario (normal aging with no mitigation) for both Array and Wallace Tree Multipliers, considering different bit widths (4, 8 12, 16 and 32-bits). From Figure 5, we see that the proposed method consistently improves lifetime across different architectures and bit-widths, achieving an average improvement of 64% compared to baseline (natural aging without any mitigation). All the lifetimes in Figure 5 are normalized with respect to the lifetime of the baseline. The improvement in lifetime happens as the Oracle intelligently decides when to apply 2's complement transformations, effectively redistributing stress away from F2F transistors. Moreover, interestingly the effectiveness of proposed method increases for larger bit width multipliers, which are more vulnerable to aging due to their higher number of F2F transistors, eventually saturating after 12 bits. As shown in Figure 5, the 12-bit Array Multiplier achieves a 75% increase in lifetime, compared to a 66% improvement in the 4-bit version. Similarly, the 12-bit Wallace Tree Multiplier sees a 60% gain in lifetime, while the 4-bit version improves by 38%. For 16-bit and 32-bit multipliers, we also see up to 62% improvement. Overall, these results highlight that the proposed technique using 2's complement is an effective mitigation for NBTI, irrespective of the multiplier architecture and bitwidths and has the potential to enhance lifetime significantly.

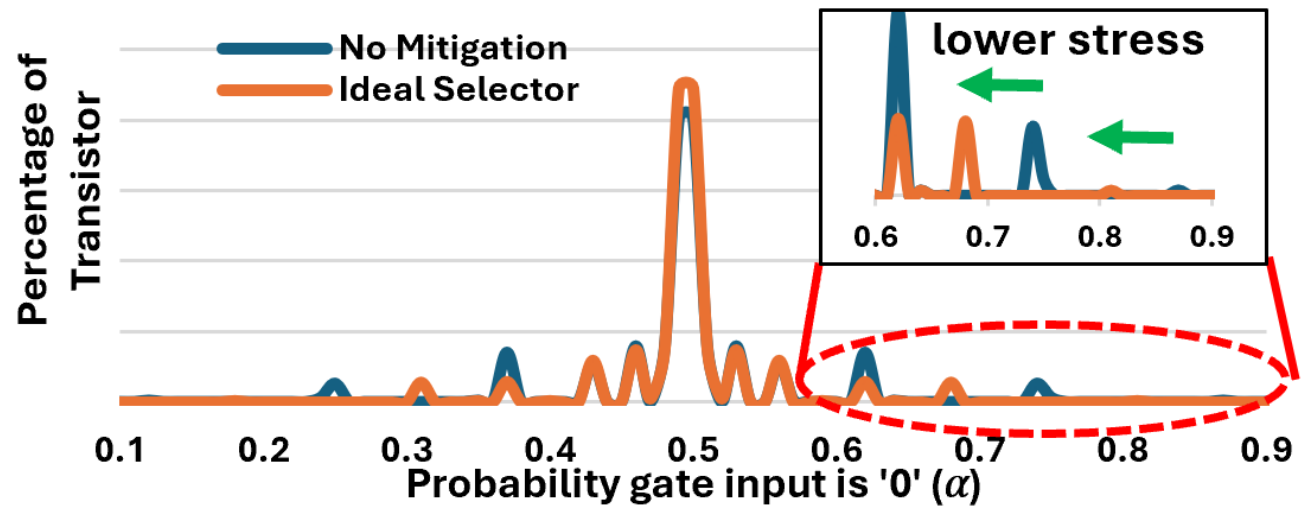

**Figure 4.** Stress distribution quantified as the probability of PMOS gate input being '0' with and without using proposed method.

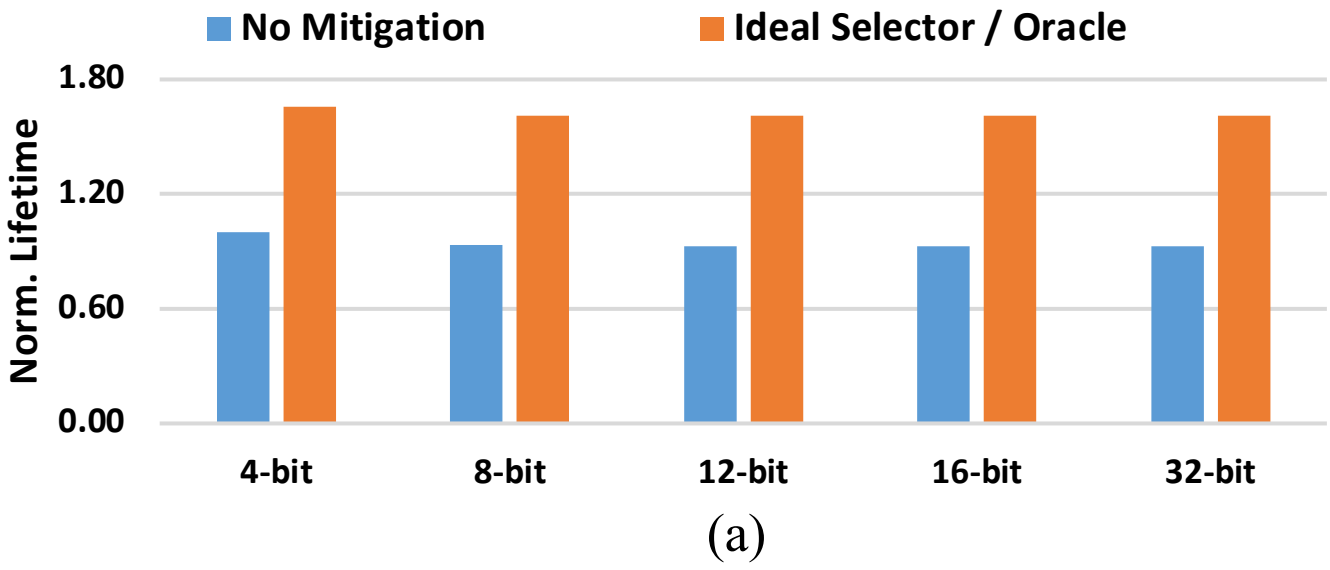

(a)

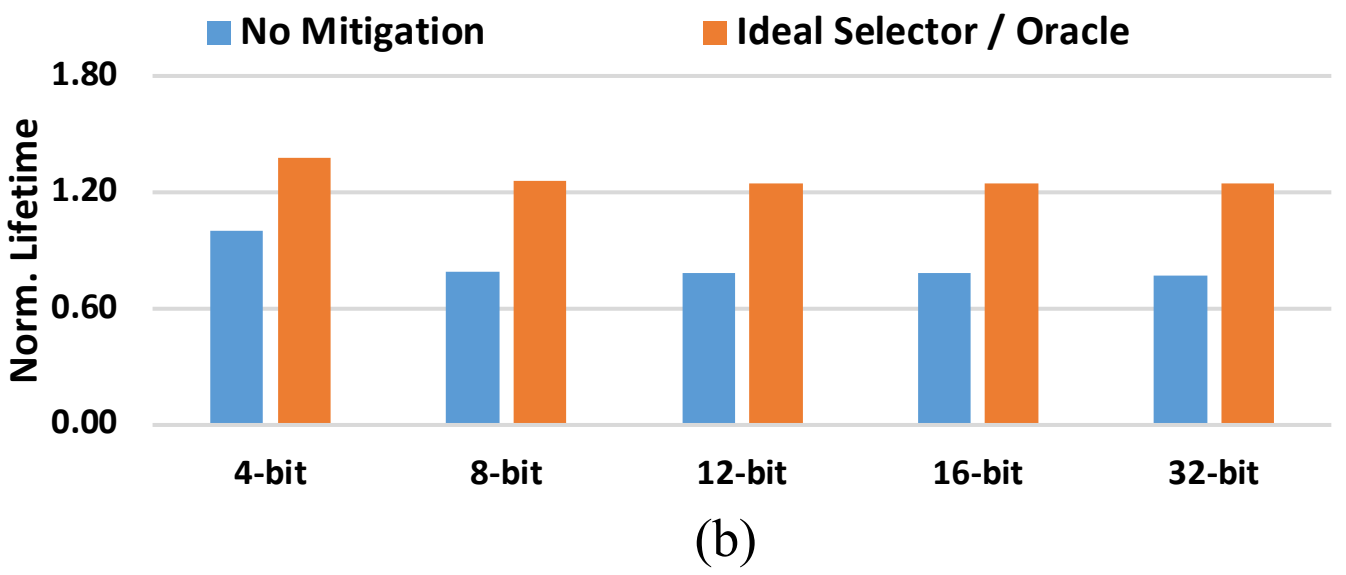

(b)

**Figure 5.** Improvement in SA lifetime (normalized with respect to the expected lifetime of 4-bit configurations) using proposed 2's complement transformation when applied using ideal selector / Oracle for (a) SA with Array and (b) SA with Wallace Tree Multiplier Architectures with different bit-widths (4-bit, 8-bit,12-bit, 16-bit, and 32-bit).

### *C. Lifetime improvement using hardware-based selector*

As discussed earlier, despite its promising results, the Oracle is not a good solution in practice due to its large overhead when implemented in hardware. Recall that for a $m$-bit multiplier, the Oracle will include $2^{2m}$ entries; Implementing it with a look-up table or a custom digital logic is expensive. Hence, we propose to develop the hardware-based selector that *approximates* the Oracle closely as discussed in Section IV. For this experiment, we do not consider any process variation yet and hence we only have a single set of F2F transistor. We show results with process variation in a later sub-section.

Figure 6 shows how closely we can approximate the Oracle using just $n \leq 2m$ bits in an $m$-bit multiplier from the input pairs $I \in (A, B)$. We quantify this via F1-score, a common metric used to evaluate how accurately a classification model performs. Figure 6 also highlights the lifetime improvement achieved using the proposed hardware-friendly SM. Here we show results for $m = 8$ bit Array and Wallace Tree multipliers noting that similar observations are made for other bit widths. In each case, we denote the obtained Boolean expression and its corresponding hardware implementation as $SM - 1$ (selector module 1).

As we can see from Figure 6, the F-1 score is expectedly low when we use fewer input bits to approximate the Oracle, but it increases as more bits are considered. For both Array and Wallace Tree Multiplier, we find that by carefully selecting just $4 - 6$ bits, we can closely approximate the Oracle (which includes 16 bits in total). Using additional bits results in no meaningful improvement in the F-1 score or the lifetime. For the 8-bit Array Multiplier using only $k = 4$ input bits, $SM - 1$ achieves an F1-score of 99.10% i.e., $SM - 1$ output matches the Oracle very closely. Similarly, for the 8-bit Wallace Tree Multiplier and with $k = 6$ input bits the $SM - 1$ reaches an F1-score of 98.44%. The use of fewer bits enables us to use a simpler Boolean expression, which can be enabled by a simple hardware design for the SM, keeping overheads low.

Upon comparing the $SM - 1$ with the Oracle ($\varphi$), we find that $SM - 1$ delivers a 72% lifetime improvement in 8-bit Array Multiplier, which is very close to the 73% improvement provided by the Oracle ($\varphi$). Likewise, $SM - 1$ achieves a 57% improvement in 8-bit Wallace Tree Multiplier, compared to 59% from the Oracle ($\varphi$). These results confirm that the approximate function and its hardware implementation $SM - 1$ is sufficient to provide an efficient trade-off between implementation cost and lifetime extension by applying 2's

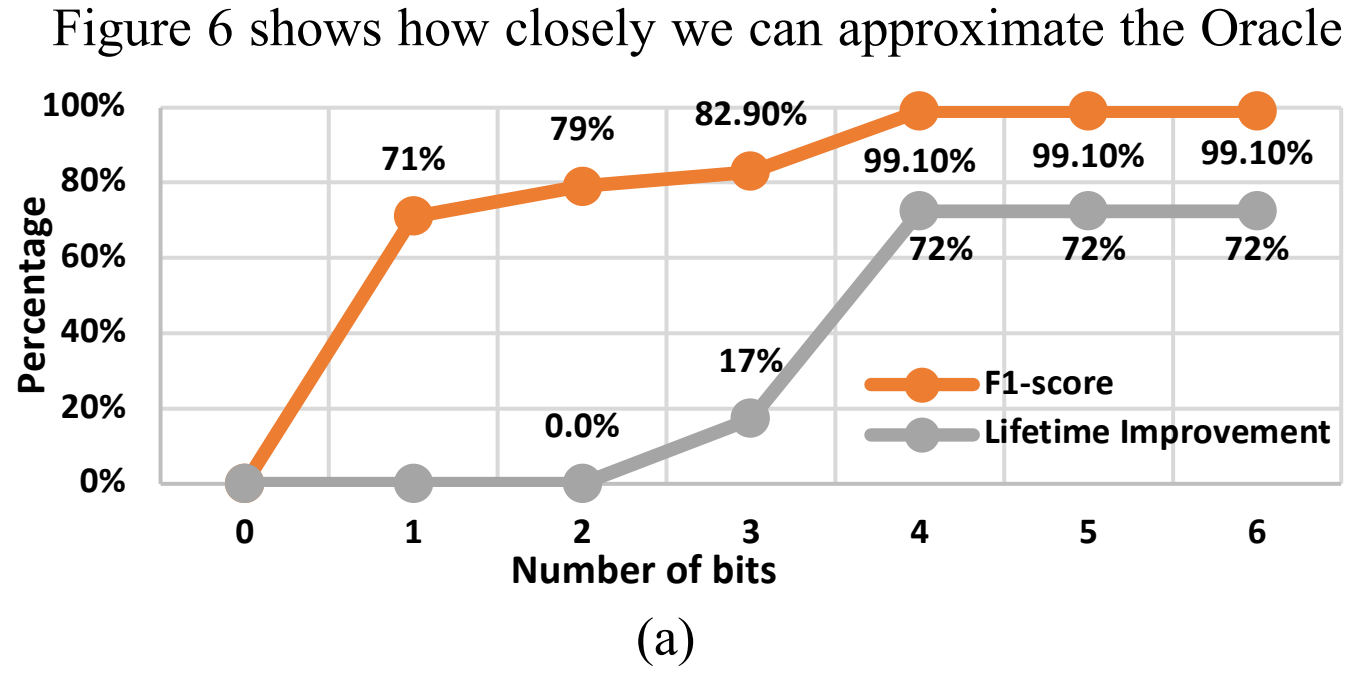

(a)

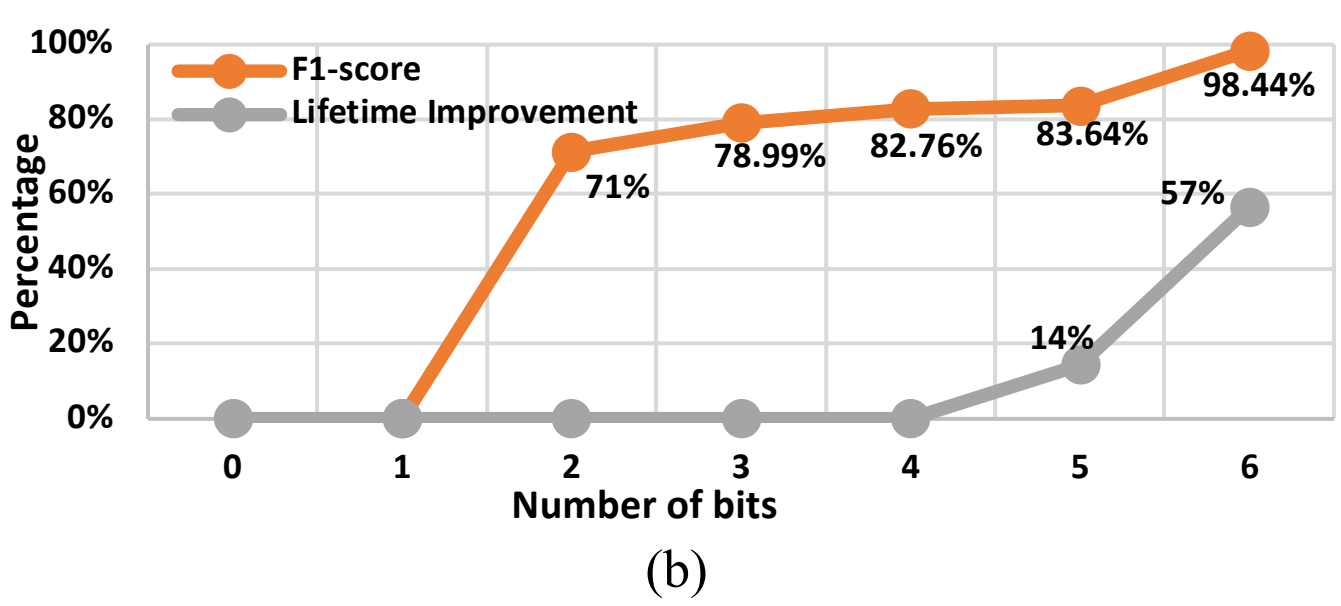

(b)

**Figure 6.** Accuracy of function approximating the Oracle ($SM - 1$) and its corresponding improvement in lifetime for (a) 8-bit array multiplier and (b) 8-bit Wallace Tree Multiplier performance.

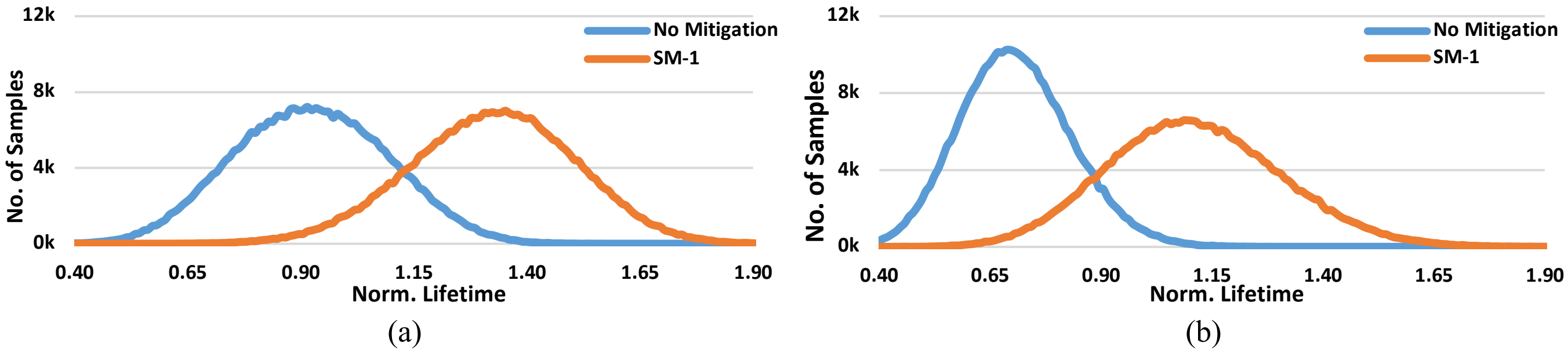

**Figure 7.** Lifetime improvement using $SM-1$ under process variation and aging for (a) 8-bit SA with Array and (b) 8-bit SA with Wallace tree multiplier architectures, based on Monte-Carlo experiment. The lifetimes are normalized with respect to the lifetime of the 4-bit SA with Array Multiplier without any mitigation, to maintain consistency with the results in Figure 5

complement selectively.

*D. Lifetime increase under joint process variation and aging*

In this sub-section, we present results in presence of both process variation and aging. For evaluating the proposed method, we performed Monte Carlo simulation considering various aging across all transistors and averaged the results to estimate expected device lifetime for a 128×128 Systolic Array (SA) architecture. In each iteration of the Monte Carlo simulation, we set initial $V_{th}$ for all transistors in the SA following Equation (4). Next, assuming a randomly chosen input to the SA in each iteration, we evaluate each multiplier. We define the lifetime of SA as the time when at least one of its multipliers has failed. As mentioned earlier, we consider failure as the point in time when the current in at least one transistor in the multiplier reduces to 50% of its unaged counterpart.

Figure 7 illustrates the expected lifetime distributions for an 8-bit SA with Array Multiplier and Wallace tree Multipliers (normalized with respect to the same baseline used in Figure 5), both before and after applying the 2's complement transformation. For this experiment, we assume the use of the same $SM-1$ from Figure 6 (prior sub-section). The distribution in Figure 7 demonstrates that the proposed method is still effective in presence of process variation as well.

As we can see from Figure 7(a), the 8-bit SA with Array Multiplier has a normalized average lifetime of 0.92, when there is no aging mitigation in place. The use of the proposed 2's complement approach (with $SM-1$) balances stress and slows aging, leading to an increase in the normalized average lifetime to 1.32, a ~44% improvement in average lifetime despite process variation. Similar observations from Figure 7(b) are made with the 8-bit SA with Wallace Tree Multiplier, which has a normalized average lifetime of 0.71 when no mitigation is used. The use of the proposed method improves normalized average lifetime to 1.11, a ~57% improvement. Note that all numbers here are normalized with respect to the lifetime of the 4-bit SA with Array Multiplier without any mitigation (to maintain consistency with Figure 5). These results show that the proposed method, can reduce stress and increase device lifetime, even in the presence of process variation.

*E. Multiple selector modules to handle process variation*

While the use of a single SM ($SM-1$) improves lifetime significantly (as we have shown in Figure 7), it does not perform well in some process variation scenarios (represented by the left tail of the bell curve in Figure 7(a) and Figure 7(b)). Adding more SMs can potentially handle diverse aging scenarios and improve circuit lifetime where each SM is tailored for a different set of F2F transistor(s). As mentioned earlier, the idea is similar to the use of ensembles in machine learning to obtain better prediction accuracy.

Figure 8 shows the impact of using $P$ SMs (where $1 \leq P \leq 5$) on the lifetime of an 8-bit SA with Array Multiplier. As we can see from Figure 8, increasing the number of SMs shifts the distribution of expected lifetime to the right, indicating a clear improvement in average lifetime. This rightward shift implies longer lifetimes even under various process variation scenarios. Interestingly, while increasing the number of SMs does lead to improved lifetime, the benefits diminish beyond a certain point. Notably, the gain going from three SMs to four SMs is minimal, suggesting diminishing returns for additional SMs beyond this point. Adding, more SMs will only lead to higher area and power overheads. Hence, using 3-4 SMs is enough to mitigate NBTI aging and increase lifetime by ~57% on average compared to baseline (no mitigation) with process variation.

Table I presents a detailed comparison. As shown in Table I, using one SM ($SM-1$) achieves ~43% average lifetime improvement despite process variation. However, using a total of three SMs provides ~57% improvement in average lifetime, a significant 14% increase in lifetime at a relatively modest increase of 2.5% in area overhead (as we show later). However, using five SMs only reaches 57.29% (0.14% additional gain) compared to using 3 SMs, while adding more area and power overheads. Hence, we use three SMs only for all scenarios.

*F. Implementation overhead*

Next, we show the implementation overhead of the proposed method in the context of a systolic array-based AI accelerator. The method can be implemented with minimal changes to the original SA architecture design as we have shown in Figure 3.

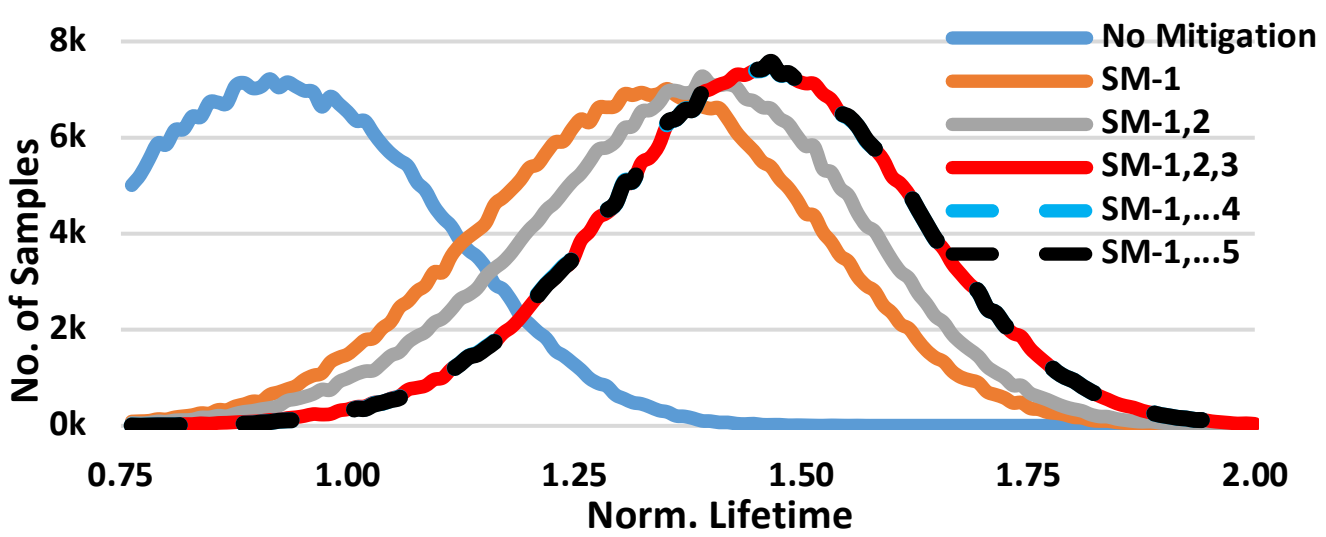

**Figure 8.** 8-bit SA with Array Multiplier lifetime using different SMs under process variation. The improvement is negligible after 3 SMs.

**Table I**
**8-bit SA with Array Multiplier lifetime using $SM$**

| Selector Module | Lifetime Improvement | Area overhead | Power overhead | Delay overhead |
|---|---|---|---|---|
| $SM-1$ | 43.57% | 1.9% | 0.17% | 2.5% |
| $SM-1{,}2$ | 48.99% | 2.8% | 0.25% | 2.5% |
| $SM-1\ldots3$ | 57.15% | 4.4% | 0.36% | 2.7% |
| $SM-1\ldots4$ | 57.25% | 6.2% | 0.49% | 2.9% |
| $SM-1\ldots5$ | 57.29% | 6.9% | 0.62% | 2.9% |

Table I shows the area overhead introduced by the proposed method as we use more SMs. Overall, from Table I, we can see that the proposed method introduces minimal (~3% on average) hardware area overhead. This happens as the SMs are approximation of the Oracle and can be implemented using a handful of minimal sized gates with 4-6 bit inputs. As such, the SM's complexity is low, typically composed of only a few basic gates. This minimalistic structure also ensures fast decision making with negligible delay and power consumption.

From a performance standpoint, the SM introduces minimal delay. For example, the $SM-1$ delay accounts for 2.5% of the computation time of the PE. The performance overhead introduced by SMs, can be completely masked if the operations are pipelined i.e., the SM computation can be done while the multiplication and addition operations in each PE in SA are ongoing, further reducing performance overhead. Table I also shows the power overhead introduce by the SMs compared to the power required for computation by each PE. As we can see, the power overhead is ~0.38% on average considering the various SMs. Overall, the proposed method introduces little area, power and performance overheads while reducing aging and increasing lifetime of SA multipliers.

### *G. Comparison with prior method*

To evaluate the effectiveness of the proposed method, we compare it with three prior works [19], [38], [39]. The method in [38] uses a true random number generator (TRNG) to randomly apply 1's complement transformation to reduce aging in memory. More specifically, the TRNG approach applies a 1's complement transformation to convert input $I \rightarrow -I$ using a probability factor $p$. We consider four different configurations of this method with $p = 0.4, 0.5, 0.6, 0.7$ (referred as TRNG ($p$) in Figure 9). The method in [39] applies 1's complement following a zero-bias probability (ZBP) such that every single bit is logic 0/1 with equal probability. As mentioned earlier, we choose these methods as baseline, since they are the closest to the proposed method. However, unlike the SM guided proposed method, both prior methods use randomness to balance the duty cycle across transistors. As we show in Figure 9, the intelligent SM-based method outperforms the random TRNG-based implementation. In contrast, the DVFS-based method in [19] regulates voltage/frequency as needed to compensate for change in transistors current due to aging and prevent failure, thus extending lifetime.

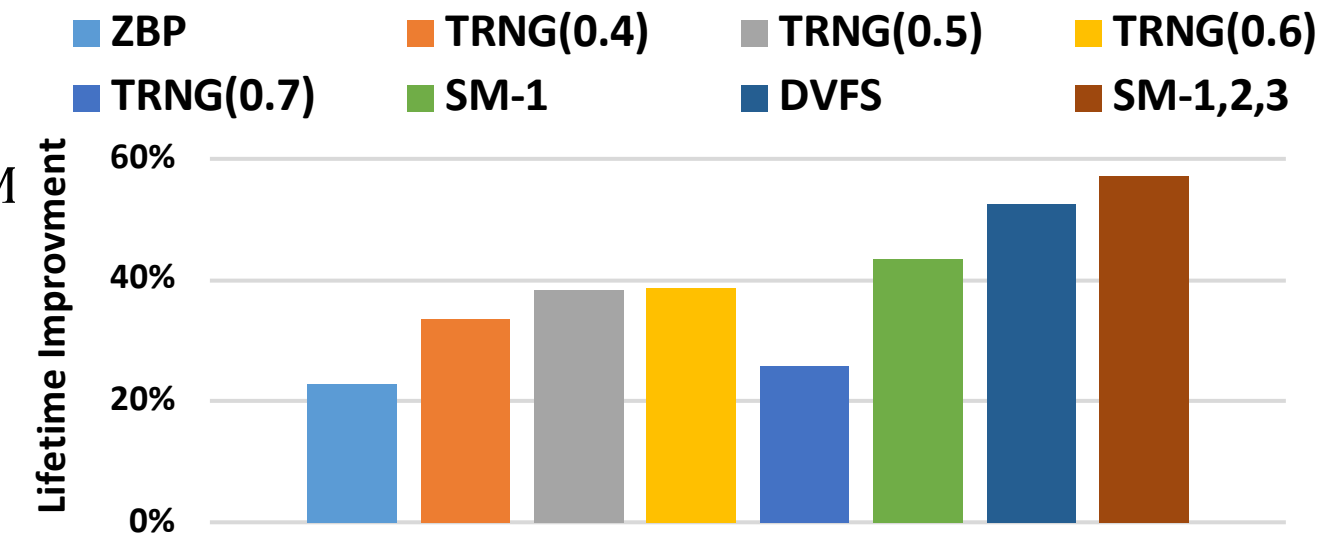

**Figure 9.** Lifetime improvement achieved by various methods compared to baseline (normal aging with no mitigation).

Figure 9 shows the comparison of all these methods in terms of usable lifetime. Interestingly, the proposed method outperforms all variants of the TRNG-based method in terms of expected lifetime by 6% even when using only a single $SM-1$. Using multiple $SM-1{,}2{,}3$ outperforms the best lifetime (at $p = 0.6$), obtained by the TRNG method, by 19%. This happens because the SM approach intelligently applies the 2's complement transformation based on inputs and the aging conditions while the TRNG approach applies the transformation randomly. Similarly, the proposed method outperforms the ZBP method by 21% using SM-1. The DVFS method improves lifetime by 53%, outperforming SM-1. However, the use of multiple SM-1,2,3, outperforms DVFS by 4.5%. From Figure 9, we see that the proposed method provides consistent lifetime improvements across different multiplier architectures, bit widths, and process variations.

### *H. Extending to Other Arithmetic Units and Types*

Thus far, we have demonstrated the proposed method for integer and fixed-point multipliers. The same principles can be extended to floating-point (FP) multipliers and also adapted for other arithmetic units, such as adders, as discussed below.

**Floating-point (FP) multipliers**: Here, we discuss the implementation assuming an IEEE 754 FP multiplier, noting that the same principles apply to other bit widths (e.g., 8-bit, 16-bit 64-bit). In FP format, a number has three distinct fields, sign, exponent, and fraction (mantissa) bits. The multiplication involves XOR-ing the sign bits, adding the exponents and multiplying the mantissa. *The mantissa multiplication is essentially an integer multiplication*, which makes it a natural target for integration with the proposed method. However, unlike fixed-point or integers, FP mantissa multiplications do not inherently support 2's complement representation for negative values. Thus, implementing the method in FP multipliers requires an additional 2's complement unit. Assuming an n-bit array multiplier, which includes $n \times (n-1)$ full adders and other logic gates, the addition of 2's complement unit ($n$-bit adder and peripherals) introduces an overhead proportional to $\sim 1/n$. This overhead diminishes with bit width; for example, in a 32-bit FP multiplier (with 23-bit mantissa), the added area is ~4%. The overhead can be reduced further via circuit optimizations.

**Extending to other arithmetic units:** The idea of applying arithmetic transformations to redistribute stress can extend beyond multiplication. While addition /subtraction are not sign-invariant, other arithmetic properties such as commutativity can be exploited. For instance, by interchanging operands $A \leftrightarrow B$, the stress is redistributed yet the sum remains unchanged as $A + B \equiv B + A$. Such transformations can be leveraged analogous to the 2's complement method for multipliers. A detailed exploration of both these extensions is left as future work.

## VI. CONCLUSION

Multipliers are one of the most ubiquitous and heavily used hardware, and as a result, are vulnerable to accelerated aging. While various aging mitigation techniques exist, they often introduce software changes or require precise scheduling and timing, which can be difficult to implement in practice. This paper presents a new aging mitigation technique that leverages the sign-invariance property of multiplications. By selectively applying 2's complement transformations via an SM the proposed method redistributes stress among transistors. As a result, it reduces stress from heavily degraded transistors and achieves significant improvement in lifetime. Due to its relatively simple implementation, it does not introduce high overheads. The proposed method has ~4%, ~0.4% area and power overheads respectively. These advantages make it a promising candidate for integration into next-generation processing platforms that support arithmetic operations such as CPUs, GPUs, and AI accelerators like systolic arrays, etc.